\documentclass[referee]{aa} 
\usepackage[T1]{fontenc}
\usepackage{amsmath}

\usepackage{graphicx}
\usepackage{txfonts}
\usepackage{natbib} 
\bibpunct{(}{)}{;}{a}{}{,} 

\newcommand{\ve}[1]{{\bf #1}}

\begin{document}

\title{Towards a theory of extremely intermittent pulsars}
\subtitle{II: Asteroids at a close distance}

   \author{F. Mottez \inst{1} \and S. Bonazzola \inst{1} 
          \and
          J. Heyvaerts \inst{2,1}\fnmsep
          }


   \institute{LUTH, Observatoire de Paris, CNRS, Universit\'e Paris Diderot,
              5 place Jules Janssen, 92190 Meudon, France.\\
              \email{fabrice.mottez@obspm.fr}
         \and
             Universit\'e de Strasbourg\\
             \email{jean.heyvaerts@astro.unistra.fr}}


 
  \abstract 
   {}
   {We investigate whether there may be one or many companions orbiting at 
   close distance to the light cylinder around the extremely intermittent pulsars PSR B1931+24 and PSR J1841-0500. These pulsars, behaving in a standard way when they are active, also ``switch off'' for durations of several days, during which their magnetospheric activity is interrupted or reduced.}
   {We constrained our analysis on eight fundamental properties of PSR B1931+24 that summarise the observations.  
   We considered that the  disruption/activation of the magnetospheric activity would be caused
    by the direct interaction of the star with the Alfv\'en wings emanating from the 
    companions. 
   We also considered the recurrence period of 70 days to be
   the period of precession of the periastron of the  companions orbit. 
   We analysed in which way the time scale of the ''on/off'' pseudo-cycle would be conditioned by the  precession of the periastron and not by the orbital time scale, and we derived a set of orbital constraints that we solved. 
   We then compared the model, based on PSR 1931+24, with the known properties of PSR 1841+0500.}
   {We conclude that  PSR B1931+24 may be surrounded at a close distance to the star by a stream of small bodies 
    of kilometric or sub-kilometric sizes that could originate from the tidal disruption of a body of 
    moderate size that fell at a close distance to the neutron star 
    on an initially very eccentric orbit. 
    This scenario is also compatible  with the properties of PSR J1841-0500, 
    although the 
    properties of PSR J1841-0500 are, by now, less constrained.}
  {These results raise new questions. Why are the asteroids not yet evaporated ? What kind of interaction can explain the disruption of the magnetospheric activity ? These questions are the object of two papers in preparation that will complete the present analysis. 
  }


   \maketitle
%

\newpage
\section{Introduction}

The pulsars \object{PSR B1931+24} and \object{PSR J1841-0500} are extremely intermittent. This means that they have two regimes of radiation. 
The on regime corresponds to the normal radiation of a standard pulsar. 
The off regime consists of interruptions of radio emissions that can last for days or months. 
The observational properties of the extremely intermittent pulsar
PSR B1931+24 have been characterised in \citet{Kramer_2006}. 
They are summarised as list of eight properties 
\textbf{P1}-\textbf{P8} in \citet{Mottez_1931p24_1}  (now called paper I). They are briefly displayed in the first line of Table \ref{table_bilan_proprietes}.

One of these properties, at least for PSR B1931+24, is a quasi-periodic behaviour of about 35 days. In  paper I, it has been found that this  cannot be a consequence of a planet orbiting the neutron star with that period, but 70 =35 $\times$ 2 days  could be the period of the precession of the periastron  of something orbiting at a close distance of the neutron star. 

The questions assessed in the present paper are the following:
(1) Why would the perturbation of the pulsar  be affected on the
precession time scale (70 days) but not on the orbital scale (a few minutes) ? A possible answer is developed in section \ref{periastre_70_jours}. It results in a series of constraints on the orbital parameters. A set of possible orbits is computed. 
(2) What kind of orbiting bodies can resist tidal forces ? This is discussed in section \ref{sec_tidal_disrupt}.
(3) Would the orbiting bodies be detected with pulsar timing ? The simple answer to this question is given in section \ref{sec_time_residuals}.

But first, we set the general principle of our investigation. 

The numerical values used in the present paper are the same as those used in paper I. For convenience,
they are displayed in Table \ref{table_parametres}.

\begin{table*}
\caption{Basic estimates for PSR B1931+24.} 
\label{table_parametres} 
\centering 
\begin{tabular}{| l |l |} 
\hline
 stellar radius &  $R_*=10$ km  \\
 stellar mass  & $M_* = 1,4$  solar mass $=2.7 \times 10^{30}$ kg \\
 stellar surface magnetic field (pole) &  $B_* =  2.6 \times 10^{8}$ T  \\
 observed period  &  $P=0.813$ s \\
 observed pulsation  &  ${\Omega}_*=7.728$ s$^{-1}$\\
observed frequency shift when on  &  $\dot{\Omega}_{on} = -10.24 \times 10^{-14}$ s$^{-2}$ \\
observed frequency shift when off  &  $\dot{\Omega}_{off} = -6.78 \times 10^{-14}$  s$^{-2}$ \\
light cylinder radius & $R_{lc} = c/\Omega  = 3,879 \times 10^7$  m $= 2,6 \times 10^{-4}$  AU\\
polar cap radius &$R_{pc} =R_* (\Omega R /c)^{1/2}$  = 160  m\\
Goldreich-Julian current  &  $I_{GJ}=-2 c \pi R_{pc}^2 \epsilon_0 \Omega_* B_*=8.6 \times 10^{11}$ A \\
Kramer current  &  $\Delta I_{pc} = 8\times 10^{11}$ A\\
\hline 
\end{tabular}
\end{table*}

\section{General principle} \label{general_principle}
To assess the first question, it is important to determine what kind of interaction is possible with the magnetosphere of the pulsar, according to the distance between the bodies and the star. 

The change of the slowing-down rate when the pulsar is off indicates the interruption of a Poynting flux from the magnetosphere associated to a polar cap current $\Delta I_{pc}$ similar to the Goldreich-Julian current (property \textbf{P5}).
This means that the magnetosphere is disturbed when the pulsar is 
off, which is bound to happen at close distance to the star, in the polar cap, where the Goldreich-Julian current fuels the whole magnetospheric activity.

In a plasma, the way of propagating energy  from one point (a body) to an another (the star or the polar cap just above it) without spreading it out in every direction is to guide it along a magnetic flux tube. When the plasma has a sub-Alfv\'enic velocity, as is expected in the vicinity of the light cylinder, this propagation is guided by Alfv\'en wings. 
\citet{Mottez_2011_AWW} have shown that a body in the wind of a pulsar would generate a wake in the form of a pair of Alfv\'en wings. Alfv\'en wings would also exist in the plasma closer to the star, inside the light cylinder.

Alfv\'en wings have been observed and theorised in other astrophysical situations, for instance, the Io-Jupiter interaction in the co-rotating plasma of Jupiter's magnetosphere 
\citep{Neubauer_1980,Bonfond_2010, Hess_2010_serpe}, Saturn and Enceladus \citep{Pontius_2006}, hot-Jupiters \citep{Preusse_2006}, planets around white dwarfs \citep{Willes_2005}, and occasionally the Earth \citep{Chane_2012}.

In some cases, the Alfv\'en wings from an orbiting body connect to the planet/star around which they turn. This is the case, in the above examples, of Io and Jupiter, Saturn and Enceladus, and it is suspected for hot Jupiters and for planets around white dwarfs.

We explore the possibility that  a  planet orbiting PSR B1931+24, or several asteroid-like bodies 
small  enough not to be seen with the pulsar timing method, behave like unipolar inductors and generate  
Alfv\'en wings that could explain PSR B1931+24's singular behaviour. 

How can Alfv\'en wings from a body influence the magnetosphere of the pulsar ? 

If the bodies are beyond the light cylinder, no energetic signal from the bodies and propagating along the magnetic field, even at the velocity $c$, can reach the star. 
In terms of magnetic energy, and of transport of material, the bodies cannot efficiently send enough energy in the vicinity of the star to perturb the magnetosphere. Of course, the wings as well as the clouds of gas emitted from evaporation would have an influence on the pulsar wind activity, but this would not hold for anything connected to the heart of the magnetosphere activity that is inside of or close to the light cylinder. All the material coming from the bodies would be dispersed away in the opposite direction. 
 
\begin{figure}
\resizebox{\hsize}{!}{\includegraphics{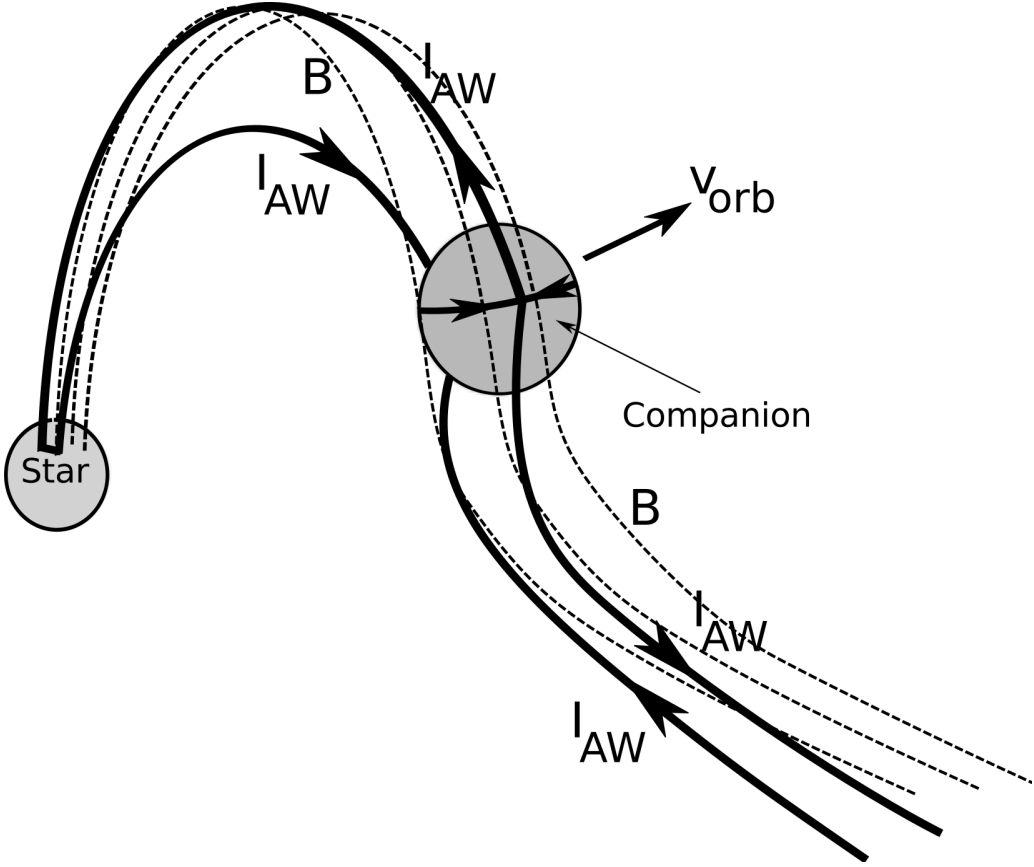}}
\caption{Thick lines represent the lines of current associated to the Alfv\'en wings of the  companion of the star,
which really form narrow ribbons rather than lines. 
They almost follow the magnetic field lines and are connected to the star, where they end. 
The currents flow into the companion
and at its surface to connect the inbound and the outbound current lines. The dotted lines represent a few 
magnetic field lines that intersect the current flowing along the companion. 
The $ \ve F=\ve I \times \ve B$ force computed by integrating over the companion's volume
has a component $F_t$ that is tangential to the direction of its motion of velocity
${\mathbf{v}}_{orb}$. The force is associated to the currents and the magnetic fields represented.}
\label{AW_dipole_shema_semi_ouvert}
\end{figure}

If the bodies are inside the light cylinder, there is a chance that the Alfv\'en wings reach the star, or a region very close to it. 
This configuration is illustrated in Fig. \ref{AW_dipole_shema_semi_ouvert}, where the companion is represented inside the light cylinder in a region of open field lines. The thin lines represent magnetic field lines. The thick lines represent the direction of the electric current carried by the Alfv\'en wings. One of these wings extends far into space, but the other is connected to the star, forming a closed system of electric currents. 
The reason why this connection is only possible
inside the light cylinder is explained in section 3.2 of paper I. Alfv\'en wings propagating
inside the pulsar's magnetosphere in relatively slow ambient plasma 
follow the magnetic field rather closely, especially when the Alfv\'en propagation speed $V_A$ (section 2.1 of paper I) is very high.

 Figure \ref{courants_AW_proche} shows the computation of the current flowing into Alfv\'en wings from bodies of various sizes. The current was deduced in \citet{Mottez_2011_AWW} in the context of two simplified models. The first corresponds to a pulsar wind (over-estimating the magnetic field amplitude and consequently the Aflv\'en wing current), the other corresponds to a co-rotating plasma in a dipolar magnetic field. In a pulsar magnetosphere, the magnetic field decreases more slowly than a dipole field, and this model under-estimates the Alfv\'en wing current. 
 Therefore, we can expect the real current to be between these pairs of values. 
 Three curves are given for each model. They correspond to bodies of  1 km, 10 km, 100 km diameter. We can expect that a few bodies of 100-km-diameter provide a total current similar to the disruption current estimated in \citet{Kramer_2006}, which is close to the Goldreich-Julian current. A similar current could be obtained with about 1000 to 10,000  bodies of diameter $\sim 1$ km. 
 
In these circumstances, the Alfv\'en wing currents can be large enough to strongly perturb the behaviour of the inner magnetosphere. 

Two opposite cases can be considered : 
\begin{enumerate}
\item \textbf{Hypothesis of a standard pulsar} The isolated pulsar is a normally active pulsar. When the bodies are outside the light cylinder, the pulsar is on. When the bodies enter the light cylinder, and especially when the Alfv\'en wings reach the star, they disrupt the magnetospheric activity. Then the pulsar is off. Because of the clouds of ionised gas evaporated from the bodies, this situation does not happen exactly when the bodies enter the light cylinder, but later, in a rather unpredictable way. 
\item \textbf{Hypothesis of a dormant pulsar} The other possibility is that the isolated pulsar is inactive, and when the bodies are inside the light cylinder, they favour conditions that enable magnetospheric activity. In that case, the pulsar would be off when the bodies are outside the light cylinder, and on when the bodies are inside it (with the same unpredictable timing caused by the evaporated plasma).
\end{enumerate}

The first hypothesis corresponds to the situation where, as for the very large majority of other pulsars, the radio waves are expected to come from inside the light cylinder. In that case, they do not result from any interaction with companions and/or Alfv\'en wings. On the contrary, the companion interrupts the normal activity of the pulsar when it is too close to the star. 

The second hypothesis is not standard, and a ``mainstream'' stance would not require that we keep it. However, a few publications have considered that some particular pulsars could be dormant unless an in-fall of neutral matter at a close distance activates the radio emissions. For instance, \citet{Luo_2007} have proposed that rotating radio transients (RRATs) could be dormant or undetectable pulsars activated by the injection and ionisation of neutral material inside the light cylinder. The proposed mechanism (requiring a relatively low neutron star rotation period) as well as the time scale of the inactive and active phase are not applicable to PSR1931+24, but it is not excluded that something similar might  happen by a different process. This is why the hypothesis of a dormant pulsar cannot be discarded a priori. Consequently, even if the hypothesis of a standard pulsar is the most probable, we kept both hypotheses in our analysis.

In either case (hypothesis of an active or dormant pulsar when not perturbed), a detailed analysis of the Alfv\'en wing interaction with the neutron star's magnetosphere is not provided in the present paper. We leave this question for the next step of our analysis in a forthcoming paper. But we retain the important fact  that the Alfv\'en wings cannot disturb the pulsar's activity through a connection to the neutron star unless the orbiting bodies are inside the light cylinder: beyond this limit, no wave slower than $c$ can propagate along the magnetic field lines down to the star or the inner magnetosphere.

As we will see in section \ref{sec_tidal_disrupt}, the bodies are not isolated, but form a stream that populates the entire orbital ellipse. Consequently, if one body is within the light cylinder, other bodies -that follow the same trajectory but with a different phase angle- are outside it. And when many bodies leave the light cylinder, others enter into it. Therefore, the perturbation of the pulsar magnetosphere is not detected on the orbital period, because it is the position of the orbit relatively to the light cylinder that counts, but not the position of each single body. Therefore, the characteristic times marking the disruption of the magnetosphere activity  is not related to the orbital period of the bodies, but to a longer time scale that characterises a precession of the orbit. As mentioned in paper I, this time scale can be the period of precession of the periastron. 

Therefore, we distinguish two classes of orbits : those that are completely outside the light cylinder, and those that are partially inside it. These
two classes of orbits determine the possibility of two different modes of activity of the pulsar. The present study focuses on the conditions permitting these two classes of orbits. We first characterise which range of semi-major axis and eccentricities allows  a period of the precession of the periastron of 70 days, and  the transition between the two classes of orbits in this course of 70 days (section \ref{periastre_70_jours}). 

Our computations are constrained by the properties \textbf{P1}-\textbf{P8} of PSR 1931+24. In section \ref{section_PSR_J1841m0500}, they are applied to PSR J1841-0500, whose properties are also summarised by a less constrained list \textbf{Q1}-\textbf{Q8} of properties. 

\section{Companions  at a short distance and a periastron precession with a period of $\sim$ 70 days} \label{periastre_70_jours}

In paper I, we reached the conclusion that the phenomena associated with
PSR B1931+24 could not be produced by a unique companion in a 35-day orbit.
We considered that the 35 days periodicity might 
be not the orbital period, but the period of precession of the periapsis of some much smaller orbit.
More precisely,we show in the following section that
the period of the precession of the periapsis should be twice as long, i.e. $P_{per} \sim 70$ days. The 
period of the precession of the periapsis is expressed in terms of the semi-major axis $a$ and the 
eccentricity $e$ of the orbit by
\begin{equation} \label{eq_gr_period_precession}
P_{per} = 
\frac{2 \pi c^2 a^{5/2} (1-e^2)}{3 (G M_*)^{3/2}},
\end{equation}
where $G$ is Newton's constant and $M_*$ the neutron star mass. 
As any property of a gravitational motion, $P_{per}$ does not depends on the mass of the pulsar's companion,
nor on its size, which is considered negligible in equation (\ref{eq_gr_period_precession}).
For a trail of asteroids orbiting along the same original orbit, the trajectory 
of each body in the trail is the same, including the periapsis precession,
so that  the whole swarm keeps on sharing a common instantaneous elliptical orbit, which precesses at a common rate.

\subsection{Geometries suitable for active and passive states} \label{sec_periastron_geometry}

In this section, we assume a standard pulsar that is
perturbed when the bodies enter the light cylinder. We could equally well assume a dormant pulsar activated when the bodies enter the light cylinder; 
we would just have to exchange the words ``off'' and ``on'', and ``active'' and ``inactive''. 
 
Since the model constrains the periastron precession period
to fit the observed quasi-periodicity, the semi-major axis $a$ is 
a function of the eccentricity $e$. Equation (\ref{eq_gr_period_precession})
can be written as follows, where $C$ is a constant known from observations, assuming that they
provide the values of $\Omega_*$ and $P_{per}$ and that $M_*$ is known:
\begin{eqnarray} \label{eq_a_de_C_e}
a &=& \frac{R_{lc}}{C (1-e^2)^{2/5}}, \\
C &=&\left( \frac{2 \pi c^2}{3}\right)^{2/5} 
\frac{R_{lc}}{(G M_*)^{3/5}}
\frac{1}{P_{per}^{2/5}}.
\end{eqnarray}
For PSR 1931+24, with a periastron drift period of 70 days, $C=0.43$.
The semi-major axis as well as the apoastron and periastron distances to the star
are plotted for this object as a function of $e$
in Figure \ref{a_e_proche_1931}.
It can be seen that, given the limitations on the eccentricity values derived below, the range of possible variations of $a$ is very narrow for PSR 1931+24.
In the following numerical estimates, we take $a=10^8$ m.

\begin{figure}
\resizebox{\hsize}{!}{\includegraphics{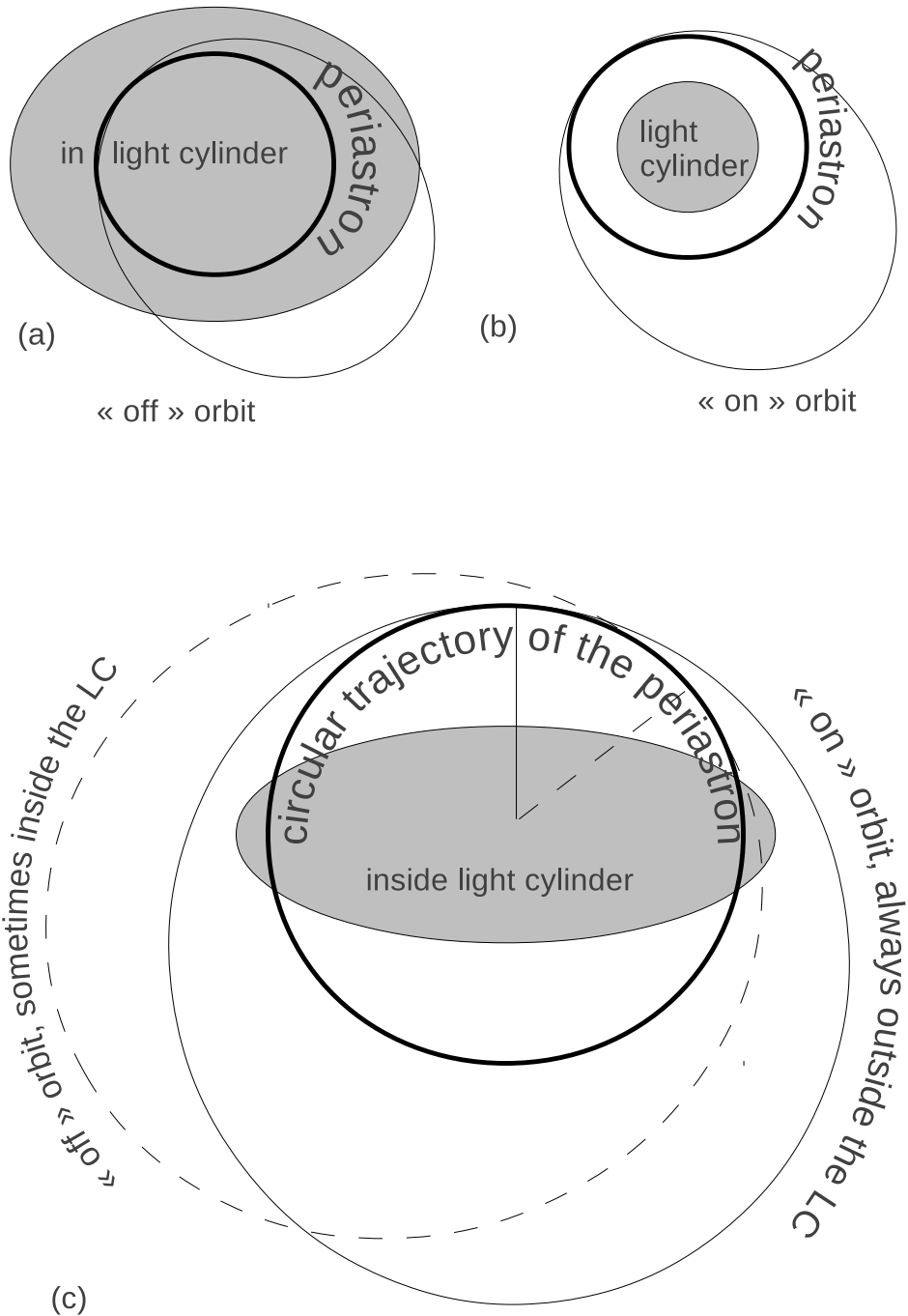}}
\caption{Sketch drawn in the companion's orbital plane. The grey area is inside the light cylinder (LC). The thick line is the circular trajectory followed (in $\sim$ 70 days) by the periastron. The thin lines are examples of orbits, approximated here by an ellipse. 
(a) The orbital plane makes a finite angle with the star's equatorial plane, and the periastron is lower than the light cylinder's radius. (b) The orbital plane is the same as the star's equatorial plane, and the periastron is higher than the light cylinder's radius. (c) An example of an inclined orbital plane and a periastron higher than the light cylinder's radius. The geometry displayed here allows for both "active" and "silent" modes.
We drew the sketch accordingly to the ``standard pulsar hypothesis'' (hypothesis 1 in section \ref{general_principle}). Under the hypothesis 2 (''dormant pulsar''), the words ``on'' and ``off'' must be permuted.  }
\label{geometry_periastron_LC_orbits}
\end{figure}

Figure \ref{geometry_periastron_LC_orbits} shows a series of sketches drawn in the orbital plane of the companion(s). 
The grey-shaded ellipse (or circle) is the intersection of the light cylinder with the orbital plane. 
The light cylinder radius is $R_{lc} = c/\Omega_*$ where $\Omega_*$ is the star spin frequency. 
The thick line is the periastron trajectory (which should be covered in $\sim$ 70 days). 
It is a circle of radius $a(1-e)$.  The thin lines are examples of the companion(s) orbit, approximated by an ellipse.
The inclination of the orbital plane upon the star equatorial plane is $i$. 
The semi-major axis of the grey-shaded area 
is $R_{lc}/\cos i$,
while the semi-minor axis 
is $R_{lc}$.

We drew the sketch accordingly to the ``standard pulsar hypothesis'' (hypothesis 1 in section \ref{general_principle}). Under the hypothesis 2 (''dormant pulsar''), the words ``on'' and ``off'' must be permuted. 

In panel (a) of figure \ref{geometry_periastron_LC_orbits}, the periastron of the companions' orbit 
is inside the light cylinder ($a(1-e) < R_{lc}$) and all orbits 
are off because all of them are, partly or totally, inside the light cylinder, whatever the periapsis phase.
A necessary, but not sufficient, condition for both on and off phases is that the periastron be more distant
from the star than the light cylinder radius $R_{lc}$.
In panel (b) of figure \ref{geometry_periastron_LC_orbits}, for example, the periastron's distance is 
larger than the light cylinder radius ($a(1-e) > R_{lc}$), but the orbit is not inclined upon the 
star's equatorial plane ($i=0$), 
and all orbits are on, whatever the periastron's phase.
The two modes alternate only
when (1) the periastron is more distant than the light cylinder radius and (2) some orbits nevertheless
cross the light cylinder at some phases of the periapsis precession.
This is only possible when the orbital plane makes a finite angle 
$i \ne 0$ with the star's equatorial plane, as in panel (c)
of figure \ref{geometry_periastron_LC_orbits}. 
A necessary, but still not sufficient, condition to have both types of trajectories
then is that the periastron distance
$a(1-e)$ be comprised between the grey ellipse's 
semi-minor and  semi-major axis:
\begin{equation} \label{eq_geometry_condition1}
R_{lc} < a(1-e) < \frac{R_{lc}}{\cos i}.
\end{equation}
When $a(1-e)$ is smaller than $R_{lc}$, all orbits are off because they either intersect the light cylinder
or are entirely included in it. 
The inverse condition $R_{lc} <  a(1-e)$ only guarantees that some orbits are partly out of the
light cylinder, but not that some are entirely external to it.
When $R_{lc}/\cos i < a(1-e)$,\, all points on the orbit are farther away from the
star than the largest distance to the light cylinder,
and the regime is permanently on. 
The inverse condition $a(1-e)< R_{lc}/\!\cos i$ 
only guarantees that some orbits (for instance for $\phi_0=0$ in Fig. \ref{deux_ellipses})
intersect the light cylinder and therefore are off, but it does not
provide any information on the existence of on states.
From equation (\ref{eq_a_de_C_e}), the inequalities in Eq. (\ref{eq_geometry_condition1}) can be written as
\begin{equation} \label{ineq_condition1_e_C}
\frac{(1-e)}{(1-e^2)^{2/5}} \,  \cos i \leq C \leq \frac{(1-e)}{(1-e^2)^{2/5}},
\end{equation}
or equivalently as:
\begin{equation} \label{ineq_condition1bis}
 e < e_{max} \; \mbox{ and } \; 
 i > i_{min}(e),  
 \end{equation}
where $e_{max}$ is defined by the equality case in the right-hand side of 
Eq. (\ref{ineq_condition1_e_C}) and corresponds to the equality 
between the periastron distance and the light cylinder radius:
\begin{equation} \label{definition_e_max}
\frac{(1-e_{max})}{(1-e_{max}^2)^{2/5}} = C.
\end{equation}
For PSR 1931+24,  $e_{max} \sim 0.65$.
The minimum inclination $i_{min}(e)$ is defined by the equality case in the left-hand side of 
Eq. (\ref{ineq_condition1_e_C}), namely:
\begin{equation} \label{definition_i_min}
i_{min}(e)= \arccos \left( C \ \frac{(1-e^2)^{2/5}}{1-e} \right).
\end{equation}

It remains to be determined, under the constraints set by Eq. (\ref{eq_geometry_condition1})
(from which the existence of off states is granted), which 
parameters $a$, $i$, and $e$ allow orbits entirely external to the light cylinder to exist, that is,
which parameters allow for both on and off regimes.
Figure \ref{deux_ellipses} shows in the orbital plane of inclination $i$ 
the trajectory of a companion at a given phase of the periapsis precession. It also shows 
the intersection of the light cylinder with the companion's orbital plane, which is the grey-shaded ellipse
with semi-major and semi-minor axes equal to $R_{lc}/\!\cos i$ and $R_{lc}$ respectively.
The polar angle of the periapsis is $\phi_0$.
The star, at $F$, is both the centre of the light cylinder's projection and the focus of the companion's orbit. 
The position $A$ of a companion on its orbit is characterized by polar angle $\phi$. 
The point on the light cylinder with the same polar angle is $C$. 
An orbit is on when for any value of $\phi$ we have $FA > FC$.
If not, it is off. The condition for $FA$ to be larger than $FC$ can be algebraically written as
\begin{equation} \label{FAplusgrandFC}
\frac{a (1-e^2) }{1 + e \cos (\phi - \phi_0)} > \frac{R_{lc} }{\cos i} \ \left(\cos^2 \phi + \cos^2 i \,  \sin^2 \phi\right)^{1/2}\, .
\end{equation}

A sufficient condition for on orbits to exist
under the conditions defined by Eq. (\ref{eq_geometry_condition1}) is that the orbit  with
periapsis phase $\pi/2$ be entirely external to the light cylinder. We establish
the validity of this statement in Appendix \ref{appsurpisurdeux}.

The condition (\ref{FAplusgrandFC}) is satisfied at this particular periapsis phase if
\begin{equation} \label{ineq_condition_2_ellipses_phi0}
[1+ e \sin \phi ] (1-  \sin^2 i \sin^2 \phi)^{1/2} < \frac{a (1 - e^2)\cos i}{R_{lc}} \, .
\end{equation} 
To find the conditions under which the inequality in
Eq. (\ref{ineq_condition_2_ellipses_phi0}) is satisfied, it suffices to compare 
the maximum value for $x \in [0,1]$ 
of the function $H$ defined by
\begin{equation} \label{definHdex}
H(x)=(1+e x) (1- x^2 \, \sin^2 i)^{1/2}
\end{equation}
to the right hand side of the inequality in Eq. (\ref{ineq_condition_2_ellipses_phi0}).
When the maximum of $H(x)$ on $[0,1]$ is lower than the right-hand side 
of Eq. (\ref{ineq_condition_2_ellipses_phi0}),
some orbits are entirely external to the light cylinder, that is on. 
We derive the condition for on and off states to both occur in a periastron precession period
in Appendix \ref{AppCNSonoff}.

When the inclination $i$ is lower than $i_{lim}(e)$ defined by Eq. (\ref{ineq_condition_1_i_petit}), 
no supplementary condition needs to be added to the inequalities in Eq. (\ref{eq_geometry_condition1})
for both on and off orbits to appear during the periastron precession period.
When $i > i_{lim}(e)$,  both on and off orbits occur
provided the inclination $i$ is lower than some upper limit $i_{max}(e)$,
defined in Appendix \ref{AppCNSonoff}. This upper limit is calculated numerically.
To summarise, when $i > i_{lim}(e)$, both on and off orbit can only exist, accounting for
the constraints in Eq. (\ref{ineq_condition1bis}),
when $i_{min}(e) < i < i_{max}(e)$ and when at the same time $e < e_{max}$.

Figure \ref{i_e_proche_1931} shows the region in the $e$--$i$ plane,
the grey-shaded area,
where both on and off states can occur depending on the periapsis phase. The constraints that apply to
cases $i < i_{lim}(e)$ and $i > i_{lim}(e)$ 
have been merged in this representation.
The grey area is bounded by the curves $i_{min}(e)$ and $ i_{max}(e)$.
For the parameters $e$ and $i$ on the left and lower side of the curve $i_{min}(e)$, 
the pulsar is always on. On the right side and for $i> i_{max}(e)$, it is always off.
The curve $i_{lim}(e)$ is plotted as well.
The division of the plane $e$--$i$ in different regions represented in Figure \ref{i_e_proche_1931}
is determined only by the value of $a/R_{lc}$, which, for a given
periapsis precession period is  determined
by the value of $C$ according to Eq. (\ref{eq_a_de_C_e}).

\begin{table*}
\caption{Numerical values used to evaluate the hypothesis of a quasi-periodicity 
induced by the precession of the periastron of PSR 1931+24. } 
\label{table_valeurs_orbite_proche} 
\centering 
\begin{tabular}{|c | c | c|} 
\hline\hline 
name & notation & value \\   \hline   \hline
 rotation period of the periapsis & $ P_{per}$  & $ 70$  days   \\  \hline
 orbital period & $ P_{orb}$  & $ 7.8$  min   \\  \hline
 semi-major axis & $ a$  & $10^5$ km = $2.5 R_{lc}$ \\  \hline
 eccentricity & $ e$  & 0 < e < 0.65 \\  \hline
 \hline
\end{tabular}
\end{table*}

\section{Why it is a series of small bodies}\label{sec_tidal_disrupt}

In this section,  we discuss the size of the neutron star's companions under the hypothesis that $P_{per}=70$ days. 
A severe constraint is imposed by the tidal forces $F_t$. 
The Roche limit $d_R$  is given in the Newtonian approximation by 
\begin{equation}
d_R \sim 2.4 R_* \left( \frac{\rho_*}{\rho_c} \right)^{1/3},
\end{equation}
where $\rho_*$ and $\rho_c$ are the neutrons star's and the companion's densities. Numerically, 
$d_R \sim 10^9$ m, which is ten times larger than the semi-major axis $a= 10^{8}$ found
to be representative
for PSR B1931+24 if the model discussed in this paper applies.
This means that a body orbiting at such a close distance cannot keep its self-gravitational cohesion.
It could  resist tearing by tidal forces however if it is small enough and if the solid material has a high enough elasticity limit $R_e$.
For iron at ambient earth temperature, the elasticity limit $R_e$,
which is homogeneous to a pressure, is $\sim$ 2.6 10$^8$ Pascals \citep{Guinier_1987}. 
A solid spherical body can resist the traction exerted by the tidal force over its whole volume,
which is about $F_t = 4G M_* M_c R_c/a_{min}^3$,
if this force is weaker than the maximum sustainable tension force $F_{lim} = \pi R_c^2 R_e$,
This allows a radius of the companion no larger than
\begin{equation}\label{elasticlimitiron}
R_{c, max} = \left(\frac{3 R_e a_{min}^3}{16 G M_* \rho_c} \right)^{1/2}.
\end{equation}  
With the ambiant's temperature value of $R_e$, this value, $R_c \sim 4$ km. 
{{The companions probably do not consist of pure iron,
even if formed from the debris of the
supernova in which the neutron star was born, nor are they
monolithic, although they probably result from the earlier dislocation
of some larger object.}}
If the asteroid is not metallic, its cohesion 
might be similar to that of terrestrial material.
Judging from the propagation velocity of shear
seismic waves in the Earth, the shear modulus of terrestrial material is about 5 10$^{10}$ Pascals \citep{Cook_1973}, and the
corresponding elasticity limit, estimated to be a lower thousand times, 
would be about 5 10$^7$ Pascals, which is fives times lower
than that of iron. Our estimate of $R_{c, max}$ in Eq. (\ref{elasticlimitiron}) 
could then be accurate to within a factor of a few.
For higher temperatures and for metallic materials, $R_e$ is less, and at 1811 K under low pressure iron melts
and its cohesion reduces to zero. It is therefore important to 
estimate the temperature of asteroids orbiting at distances compatible
with the existence of on and off orbits.
The sources of heat include the pulsar thermal radiation and the luminosity associated to the spin-down of the pulsar. The spin-down luminosity is caused by the wind of highly relativistic particles, and the very low frequency ($\omega = \Omega_*\sim 8$ Hz) and very high amplitude electromagnetic wave. In the null-frequency approximation, the spin-down radiation is the cause of the Alfv\'en wings, and these wings carry electromagnetic energy far into space, where it is probably radiated \citep{Mottez_2011_Graz}. Then, only a fraction (to be estimated) of the spin-down luminosity received by the asteroid is directly converted into thermal energy and melting. 

Thus, in spite of the high amount of radiated energy at close distance from the neutron star, it is assumed in the present paper and the following (paper III, in preparation) that the asteroids are solid. The question of their temperature and their time of fusion will be discussed in paper IV.

\section{Time residuals} \label{sec_time_residuals}
Property {\bf P8} requires that the companions cannot be detected with the method of time residuals. 
\citet{Rea_2008} 
have made an analysis of the timing residuals based on a companion with a period of 35 days ($a=900 R_{lc}$). Their analysis is based on the R\"omer delay (the light-travel time across the orbit of the star around the barycentre of the star and the companion). This delay $\Delta t_{R}$ scales as $x=a_* \sin i$, where $a_*$ is the semi-major axis of the star's motion around the barycentre, and $i$ is the inclination of the orbit to the line of sight. 
As $a_* \sim a M_c/M_*$, we can see that the Roemer delay decreases when the companion is closer to the star. 
For a companion and $a=2.5 R_{lc}$ ($\sim$ 7.8-min orbit), it is $370$ times smaller than for a planet with the same mass and $a=900 R_{lc}$ (35- days-orbit). \citet{Rea_2008} showed that with $a=900 R_{lc}$, an Earth-like companion would remain undetected. In the present case ($a=2.5 R_{lc}$) even a companion with 370 Earth masses would remain undetected with the timing residual. 
We have seen that the body at the origin of the stream of small bodies orbiting the pulsar could have had a radius 
that did not exceed 100 km. We can see there that 
even if the stream of small bodies were to remain localized on a small portion of the orbit, the criterion \textbf{P8} would be satisfied to a very large extent.

It is also possible to make a simple direct computation to confirm the above conclusion. 
Instead of considering a stream of asteroids (for instance $10^6$  asteroids with a radius of 1 km), we consider the equivalent volume concentrated in a single asteroid with a radius of $R_c=100$ km. Then, the effect on the R\"omer delay is considerably increased. The R\"omer delay for such a single body is given in \citet{Lorimer_2012}, and an overvalue can be easily obtained,
\begin{equation} \label{Romer_delay}
\Delta t_{Romer} = \frac{a_c \sin i}{c} [(\cos E -e) \sin \omega + \sin E \sqrt{1-e^2} \cos \omega] < \frac{a_c}{c},
\end{equation}
where $E$ is the eccentric anomaly, $\omega$ is the angle between the periastron and the ascending node, and $a_c = a (M_c/M_*)$ in the limit $M_c \ll M_*$.
A numerical estimate is based on the numerical values in Tables \ref{table_parametres} and \ref{table_valeurs_orbite_proche} and $R_c=100$ km and a density of the companion $\rho_c \sim 3 \times 10^{3} $ kg.m$^{-3}$. We find $a_c=0.3 \times 10^{-3}$m, and $\Delta t_{Romer} < 10^{-11}$s. This is nothing compared with the 
measured time residuals $\Delta t \sim 2 \times 10^{-3}$s, and this could not help for a detection.

\section{Is the model retained for PSR B1931+24 compatible with PSR J1841-0500 ?}\label{section_PSR_J1841m0500}

The singular properties of PSR J1841-0500 are presented in \citet{Camilo_2011} ; they are summarised below and numbered in a way that allows an easy comparison with the properties of PSR B1931+24.
Because of the more recent discovery of the peculiar behaviour of PSR J1841-0500, its properties are less well documented than those of PSR B1931+24.
\begin{itemize}
\item {\bf Q1} PSR J1841-0500 has two modes of radio emissions. It behaves like an ordinary pulsar during active (on) phases, then switches off and remains undetectable (silent/off phases). 
\item {\bf Q2} Up to now,  one long off phase (from 2010-1-19 to 2011-7-26, 580 days) between two active phases has been observed. But \citet{Camilo_2011} have also mentioned that on 2009-12-11, two weeks before its long off phase, the pulsar became inactive for between 10 and 162 minutes. The authors also mentioned an image at radio wavelengths from the MAGPIS survey \citep{Helfand_2006} where the pulsar was in the field of view and undetected. 
\item {\bf Q3} During the active phases, the pulsar slows down  at a faster rate 
than during the quiet phases. For PSR 1931+24, the ratio is 1.5, for PSR J1841-0500, it is 2.5.
\item {\bf Q4} The period of PSR J1841-0500 is $P=0.9$ s. When on or off, the pulsar is placed in the standard "second pulsar" family in the $P/\dot P$ diagram. It is far from the death line. 
\item {\bf Q5} The period decreased less during the off period than in the on regime. There could have been a large glitch during the off regime, but this is rare for such large $\dot \Omega$ \citep{Espinoza_2011}. More probably $\dot{P}_{off} < \dot{P}_{on}$. This implies that the difference between the two regimes is not a mere change of beam direction, but that there is also a change of torque. In contrast to PSR B1931+24,  the $\Delta I_{pc}$ current that would explain this difference of torque has not been estimated up to now.
\item {\bf Q6 is not documented} Before turning off for 580 days, it was observed that the pulsar 
turned off in relatively brief episodes in the year preceding the long off period.
\item {\bf Q7 is not documented} The period change between the on and off phases is not measurable.
\item {\bf Q8 is not documented} No explicit mention is made, up to now, of a companion that would be detectable with the timing residual.
\end{itemize}

These properties are similar to those of PSR 1931+24. The main difference concerning the present model is that 
the periastron drift period would be much longer. We have arbitrarily chosen a periastron drift period $P_{per}=1000$ days, which is compatible with the off period duration of 580 days. Then $C \sim 0.17$.  
We have plotted the allowed domains of inclination, eccentricity in Fig. \ref{i_e_proche_1841}. We can see that the allowed range of eccentricities is even broader than for PSR 1931+24. But the orbit inclination is globally 
higher. The semi-major axis is plotted in Fig. \ref{a_e_proche_1841} as a function of the eccentricity. For PSR 1841+0500, the expected distance to the asteroids would be larger.

Another pulsar, PSRJ 1832+0029 \citep{Lorimer_2007}, is known to behave in a similar way to PSR 1931+24 and PSR 1841+05, but its characteristics are poorly documented in the literature. The time scale of the switching between off and on states is of the order of hundreds of days, the  pulsar's spin-down rate almost doubles when the radio emission is on. Unfortunately, we do not know (the authors of the present paper) if its behaviour is quasi-periodic.


\section{Discussion and conclusion}

The analysis presented in this paper 
is based on a series of observed peculiarities of the pulsar PSR B1931+24, numbered  here \textbf{P1}--\textbf{P8}. 
We aimed to formulate a theoretical explanation that would be consistent with all these properties. 
In paper I, we assumed that the quasi-periodicity of 35 days of the behaviour of PSR 1934+21 (property \textbf{P2}) is caused by a single body orbiting the neutron star. We also considered
that the coupling process between the star, its companions, and the radio emissions is caused by 
the Alfv\'en wings that are carried by the companions when they move in a sub-Alfv\'enic plasma (the pulsar's wind, or its magnetosphere).
No model based on a body orbiting in 35 days or 70 days could explain all properties
\textbf{P1}--\textbf{P8}. Then, we considered that the body might be orbiting 
at a close distance to the light cylinder. We showed that in that case, 
the quasi-period would be a consequence of a strong precession of the periastron, that would have an overall period of 70 days. This precession would be induced by the relativistic gravitational field of the neutron star.  But this orbit cannot 
be occupied by a single big planet, because it would be disrupted by tidal forces. 
It would be occupied by a stream of small bodies, possibly of radius close to 1 km.

 This configuration could explain 
the two different regimes of the pulsar (property \textbf{P1}) if we admit that when in the light cylinder, the bodies interact with the star in a way that disrupts the activity of a normal pulsar, or, oppositely, switches on the pulsar activity of a normally dormant neutron star. 
The same principle of interaction between a pulsar and a stream of bodies orbiting 
close to it seems compatible with the observed properties of PSR J1841-0500, and a range of orbital elements allowing for the two on and off modes can also be found, as illustrated in Figs. \ref{i_e_proche_1841} and \ref{a_e_proche_1841}.
The plasma processes involved in the activation/switching-off of the pulsar activity will be discussed in paper III, in preparation.

The quasi-periodicity of the transition between on and off states (property \textbf{P2}) would derive from the period of precession of the periastron. 

\begin{figure}
\resizebox{\hsize}{!}{\includegraphics{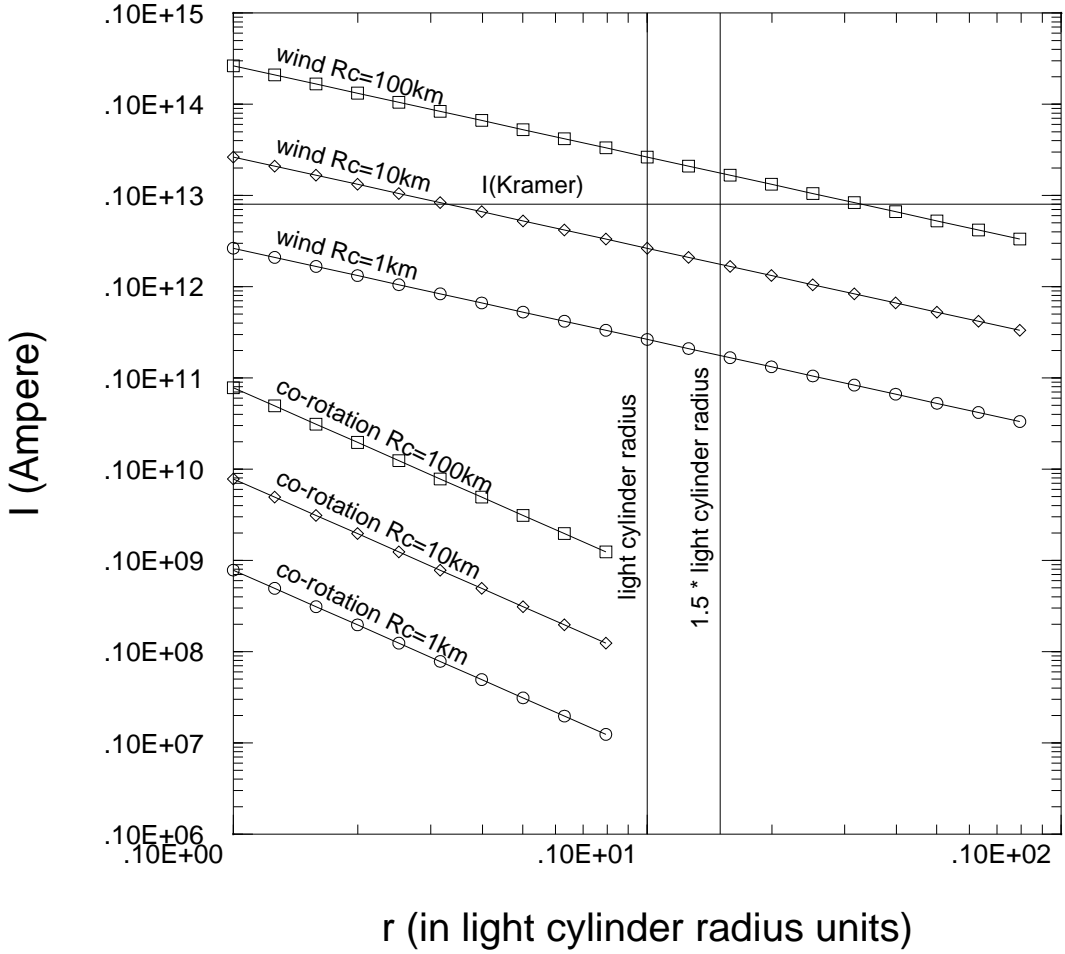}}
\caption{Dependency of the electric current carried by the Alfv\'en wings, as a function of the distance. The distance corresponding to the light cylinder radius is indicated by a vertical line. The expressions of the electric current are taken from 
\citet{Mottez_2011_AWW}. These expressions depend on the geometry and the intensity of the magnetic field. The three upper lines correspond to bodies in the pulsar wind  (with an aligned magnetic field), the three others are based on a dipole magnetic field of null inclination and a co-rotating plasma. (It is not plotted beyond the light cylinder radius, because at these distances, a co-rotating motion is not possible.). The real magnetic field and the Alfv\'en wing current inside the light cylinder are comprised in between these two approximations.  For comparison, the current computed in \citet{Kramer_2006}, characterising the pulsar property \textbf{P5}, is indicated by a horizontal line.}
\label{courants_AW_proche}
\end{figure}

The property \textbf{P3} (different slowing-down rates when on and off)
results from the hypothesis that the pulsar activity would be disrupted (we did not involve a mere directivity effect/ hiding of the radio beam). This property as well as properties \textbf{P4} (the pulsar belongs to the standard family) \textbf{P5} (the current disruption estimated from the different rates of slowing-down is similar to the Goldreich-Julian current) and \textbf{P6} (the transition from on to off is fast) depends on the model of pulsar activity disruption/activation which will be discussed in paper III.
 At the present level, we cannot conclude about these properties, but we can already see that the considerations developed in the present paper are not incompatible with them. 
Nevertheless, Fig. \ref{courants_AW_proche} shows that the Alfv\'en wing currents carried by about 1000 to 10,000 bodies with a diameter of 1 km would allow for the total current involved in property \textbf{P5}. This is not in itself a sufficient explanation, but a clue that indicates a lower estimate of the number of bodies that would orbit around the star. 

The property \textbf{P7} says that the pulsation periods are the same in the two modes. This period was directly measured in the on mode and extrapolated  in the off mode, then compared when the on mode is activated \citep{Kramer_2006}. This property is compatible with our hypothesis that the radio emissions come from the pulsar, and not from the stream of bodies orbiting it. 

The absence of timing residuals, property \textbf{P8}, is compatible with a stream of bodies of kilometric size orbiting the pulsar, as discussed in section \ref{sec_time_residuals}.

Obviously, such a stream of small bodies could originate from a larger body that fell close to the star 
and  was disrupted by tidal forces. For a body that orbits the star in the direction opposite to the star's spin, 
\cite{Mottez_2011_AWO} have shown that both the semi-major axis and the eccentricity decrease as long as the body spends most of its time in the pulsar wind.

Therefore, it is possible to consider that originally, the orbit was more eccentric and more elongated. 
This is compatible with the hypothesis of the falling of an initially distant body. 
Such orbits could be compared, in the solar system context, with those of the sun-grazing comets \citep{Bailey_1992}.

\begin{figure}
\resizebox{\hsize}{!}{\includegraphics{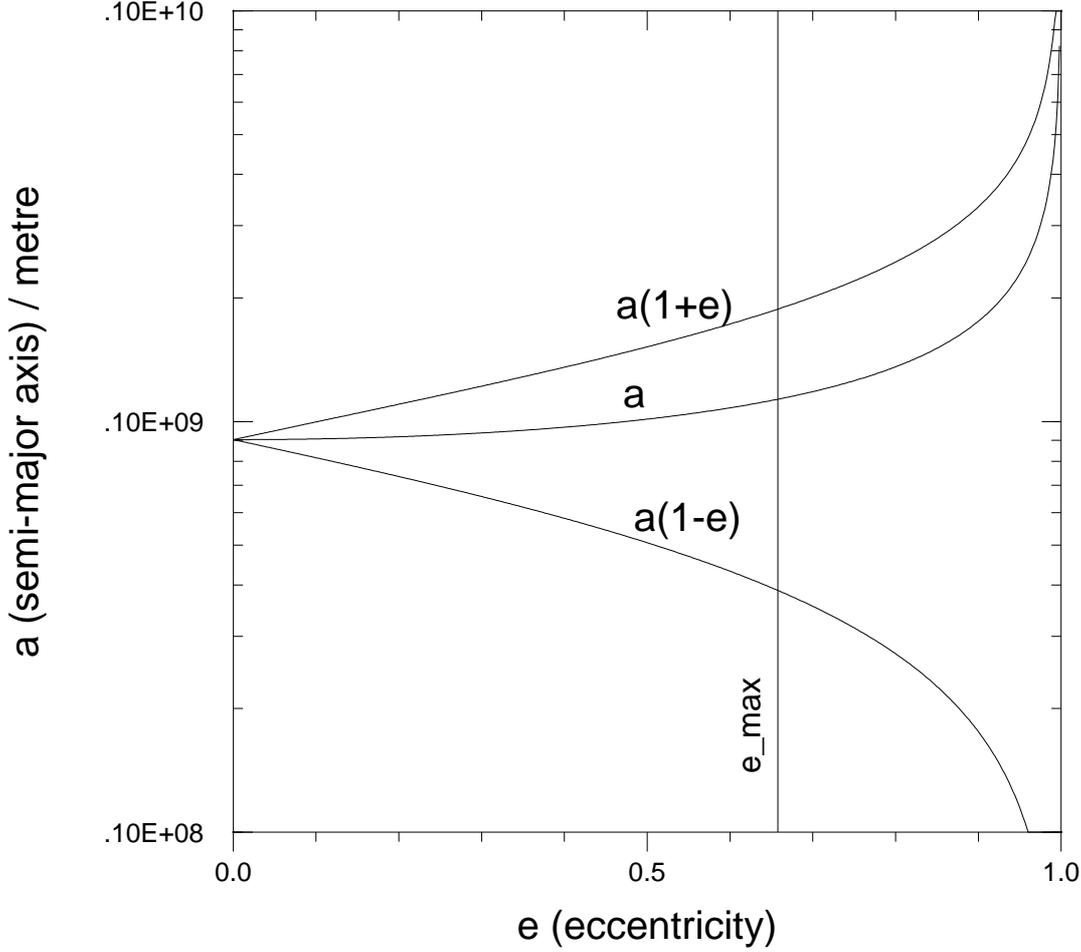}}
\caption{Semi-major axis as a function of the eccentricity for a precession of the periastron with a period of 70 days. The parameters are those infered for PSR 1931 + 24, and the relation is derived from Eq. (\ref{eq_gr_period_precession}). The two other curves represent the periastron and the apoastron.  The semi-major axis $a$ is given in units of the light cylinder radius $R_{lc}$. The values of the semi-major axis compatible with both on and off modes correspond to the part of the curve on the left-hand side of the vertical line $e=e_{max}$.
This defines a rather narrow domain of a possible semi-major axis: $0.9 \times 10^8 <a<1. \times 10^8$.}
\label{a_e_proche_1931}
\end{figure}

\begin{figure}
\resizebox{\hsize}{!}{\includegraphics{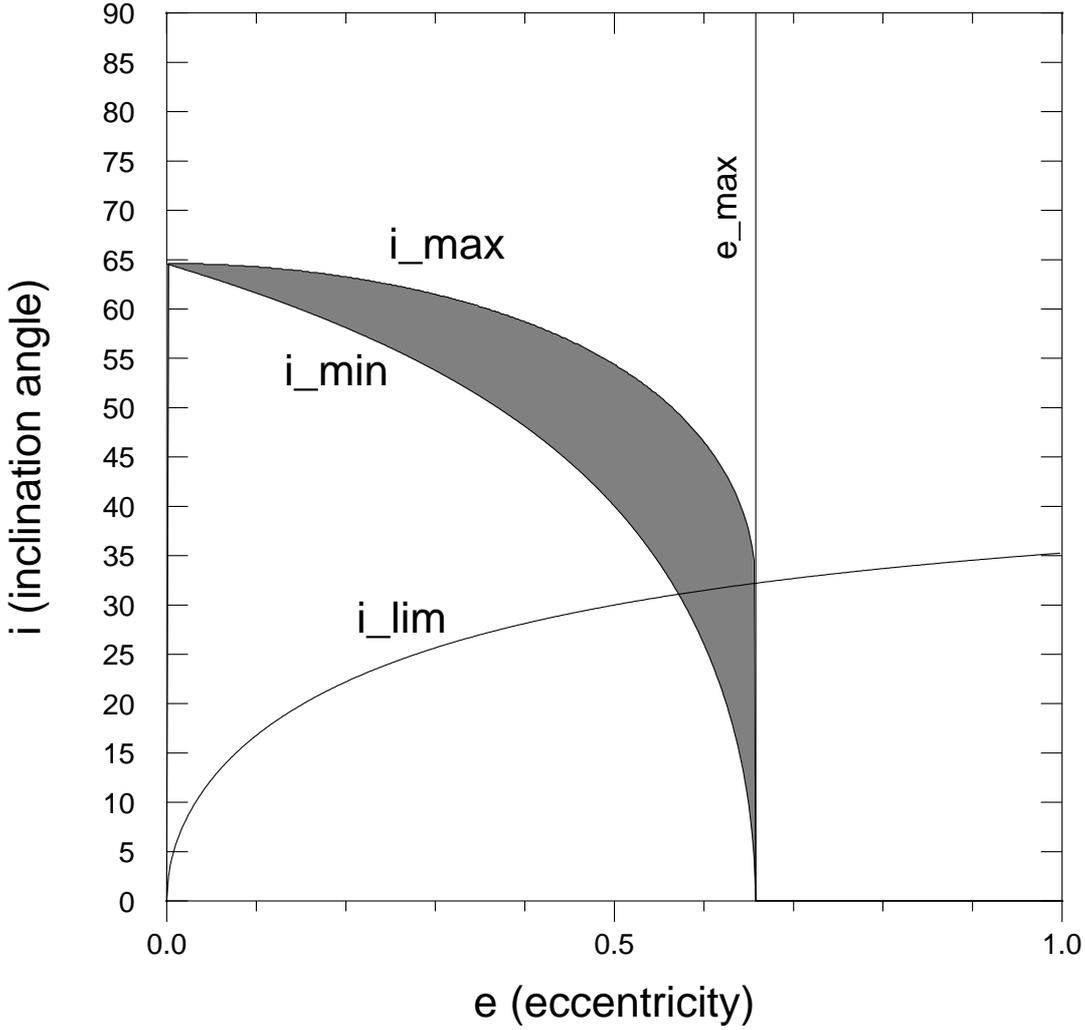}}
\caption{Lines $i_{lim}(e), i_{min}(e), i_{max}(e)$ defined 
in section \ref{sec_periastron_geometry}. The numerical values are derived for PSR1931+24.  The shaded area corresponds to the set of parameters $i$ and $e$ that allow both on and off modes. The 
vertical line at $e = e_{max}$ (eq. (\ref{definition_e_max}))
limits the domain where  on an off regimes may occur.
These exist in the shaded area. Below the curve $i=i_{lim}(e)$ (equation (\ref{ineq_condition_1_i_petit})
no additional condition applies. Parameters allowing on an off regimes in the region
$i > i_{lim}(e)$ require $i < i_{max}(e)$, the latter is defined in
the text following eq. (\ref{condisurxM}).
There is a perfect continuity between these two domains.}
\label{i_e_proche_1931}
\end{figure}

\begin{figure}
\resizebox{\hsize}{!}{\includegraphics{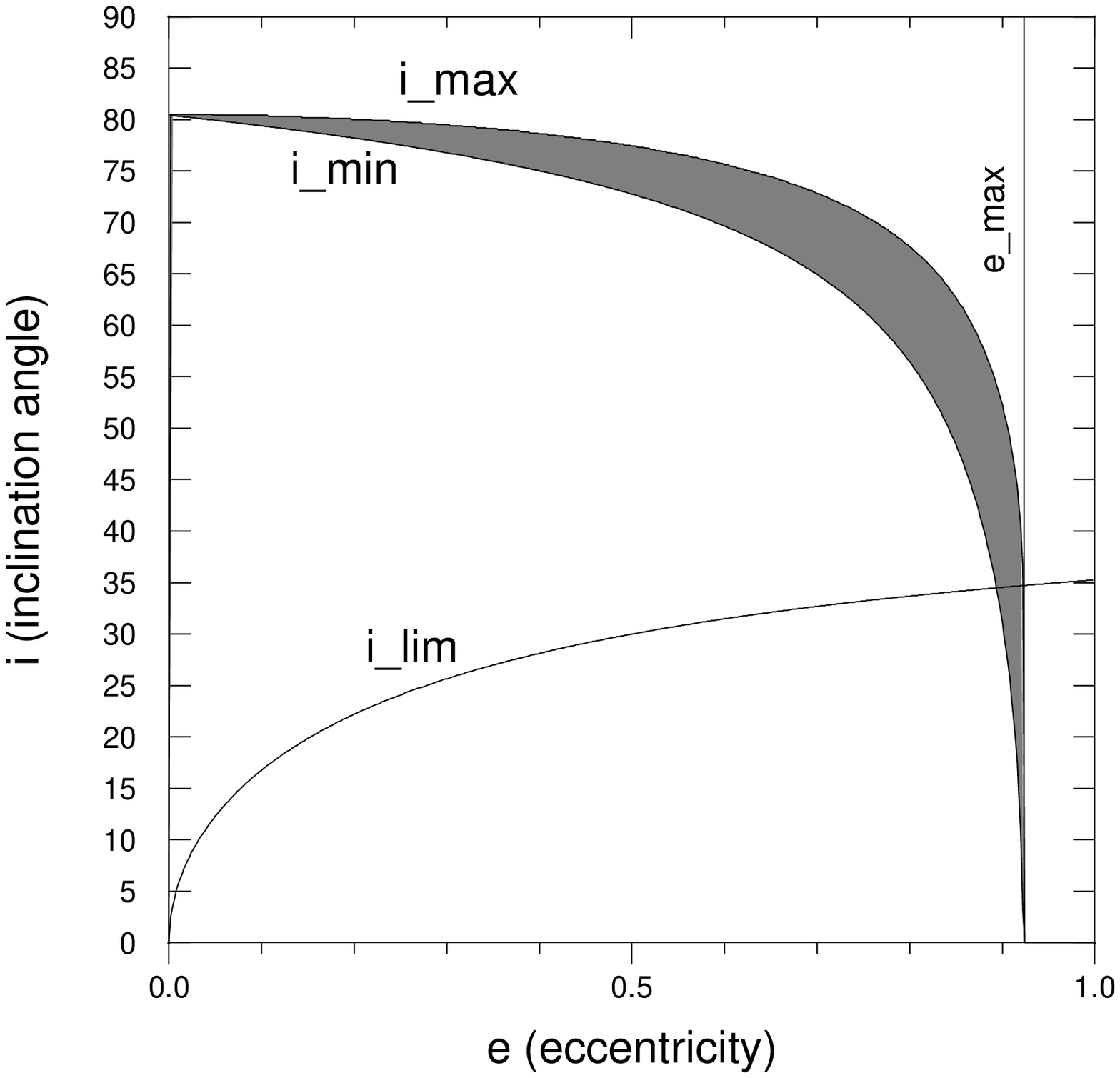}}
\caption{Same as Fig. \ref{i_e_proche_1931} for PSR 1841+0500.}
\label{i_e_proche_1841}
\end{figure}

\begin{figure}
\resizebox{\hsize}{!}{\includegraphics{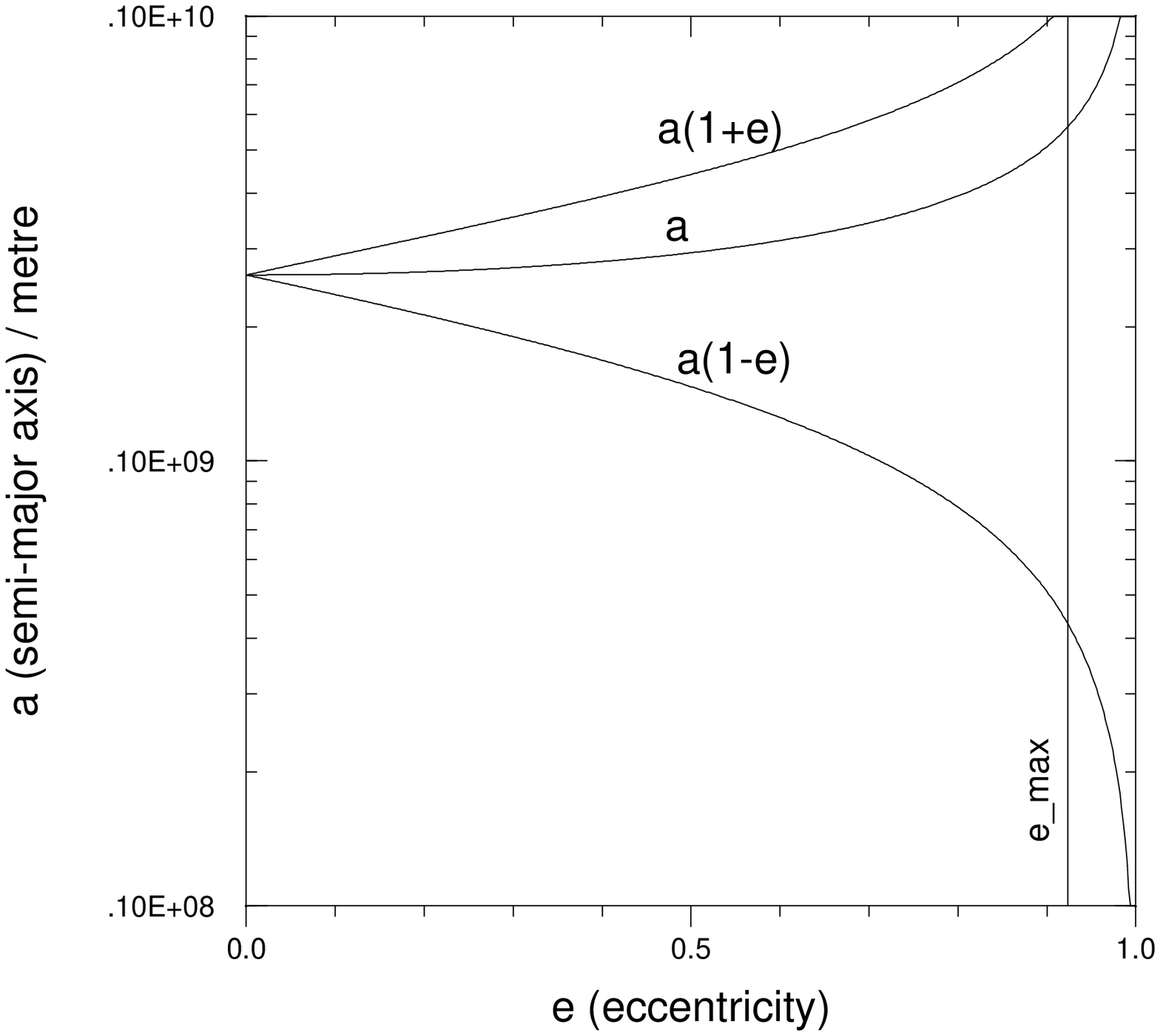}}
\caption{Same as Fig. \ref{a_e_proche_1931} for PSR 1841+0500.}
\label{a_e_proche_1841}
\end{figure}

Nevertheless, we must not conclude that any celestial body falling near a pulsar would transform 
it into an intermittent pulsar. As seen in Figs. \ref{i_e_proche_1931}, \ref{a_e_proche_1931}, 
\ref{i_e_proche_1841}, and \ref{a_e_proche_1841}, having a high eccentricity is not a sufficient criterion for long-duration off  phases. A restricted domain in the $a,e, i$ space must be reached. 
This can explain the scarcity of extremely intermittent pulsars.

Recently, \citet{Shannon_2013} have shown that another pulsar, \object{PSR B1937+21}, probably has an asteroid belt. 
 PSR 1937+21 is a millisecond pulsar and is not intermittent. The presence of an asteroid belt is suggested by the analysis of the timing residuals (favoured by the large distance of the asteroids from the star). 
 Therefore the physical contexts in PSR 1931+24 and PSR 1937+21 are somewhat different.
 Nevertheless both \citet{Shannon_2013} and us conclude on the possible existence of an asteroid belt. 
 It is important to mention that the belt around PSR 1937+21 is found to be beyond the melting distance, estimated to be 0.3 AU for a metallic asteroid. At such large distances, asteroids can be solid in their equilibrium state, which allows the asteroid belt for a very long time duration. In the case of PSR 1931+24, the asteroids are probably evaporating. This is no stable regime, but a temporary 
 situation that might end in a  short delay (maybe a few 10's of years, as will be discussed in a forthcoming paper).

Of course, the present analysis of PSR B1931+24 is based on very rough estimates. 
Our goal was to provide a {{plausible explanation of}}
the singular behaviour of PSR B1931+24 and PSR 1841+0500. {{It was not to provide}}
a detailed analysis of the characteristics of these systems, {{although we have indicated a
possible cause for each of the important aspects of the model that we outlined. 
A detailed analysis should be carried out at a later stage, developing more sophisticated models.}}
First, as mentioned, the physical process of interruption/activation of the pulsar activity must be explained. Paper III is devoted to that question. 
Two other points incompletely studied may invalidate the present model, (1) temperature, (2) orbit stability.

If the bodies are too hot, they melt, or at least their resistance $R_e$ to the traction by tidal forces is much reduced. 

The stability of the orbit defined in this model must also be investigated. 
For this a correct model of the magnetic field in the vicinity of the light cylinder need to be considered  
in regions of high latitudes, in connection with the model of the Alfv\'en wings used throughout the present work. 

Because of the temperature and stability problems, we expect that the extremely intermittent behaviour
has a finite duration, which remains to be estimated.

\begin{table*}
\caption{Summary of the different hypotheses, and of the properties that they satisfy, or fail to satisfy. } 
\label{table_bilan_proprietes} 
\centering 
\begin{tabular}{|c | c c c c c c c c|} 
\hline\hline 
  & {\bf P1}  &{\bf P2}  &{\bf P3}   &{\bf P4} &{\bf P5} &{\bf P6} &{\bf P7} & {\bf P8}\\ 
  & on/off & $\sim$ periodic & $\dot P_{off}<\dot P_{on}$ & standard & equivalent& fast & $P_{on}=P_{off}$ & no timing \\ 
  &        &                 &                            & $P \dot P$& $\Delta I_{pc}$& transition &          & residual\\ 
\hline \hline
neutral bodies & yes & no & yes & yes & yes & --& yes& yes\\
\hline \hline
$P_{orb} \sim 35$ days. &&&&&&&&\\
\hline 
circular equatorial  & yes &  no & wrong values & yes & OK... & yes & yes& yes\\
\hline 
elliptical  & yes &  yes&  wrong values & yes & ... for Earth-like & yes & no& yes\\
\hline
circular inclined  & yes & yes & wrong  values & yes & planet only. & yes & no& yes \\
\hline \hline
$P_{prec} \sim 70$ days. &&&&&&&&\\
\hline
one planet & yes & yes & -- & yes & -- & -- & yes & yes\\
 & & but & destroyed & by & tidal & forces&  & \\
\hline 
stream of bodies  & yes & yes & in paper III & in paper III & in paper III & yes & yes& yes\\
\hline \hline
\end{tabular}
\end{table*}

\newpage




\begin{figure}
\resizebox{\hsize}{!}{\includegraphics{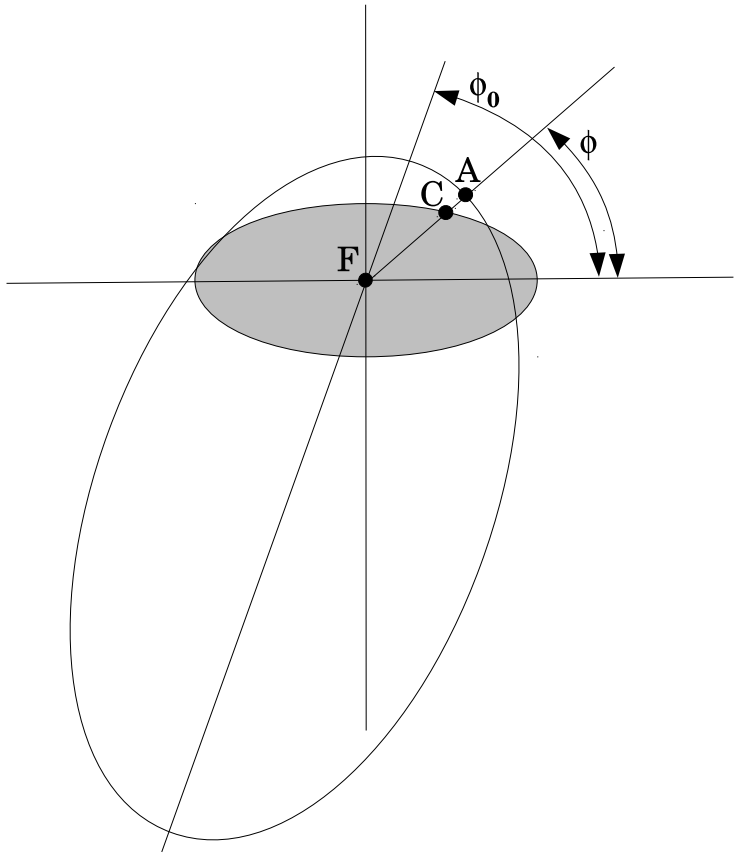}}
\caption{Orbit at a given phase $\phi_0$
of the periastron precession of a companion A is the oblique ellipse. The grey-shaded ellipse is the projection of the 
light cylinder onto the orbital plane, and C is the point on the light cylinder projection at the same phase
as A. The star F is both the focus of the orbit and the centre of the light cylinder projection.
The polar angle of the periastron is $\phi_0$, and the instantaneous polar angle 
of the companion A along its trajectory is $\phi$.
}
\label{deux_ellipses}
\end{figure}

\begin{figure}
\resizebox{\hsize}{!}{\includegraphics{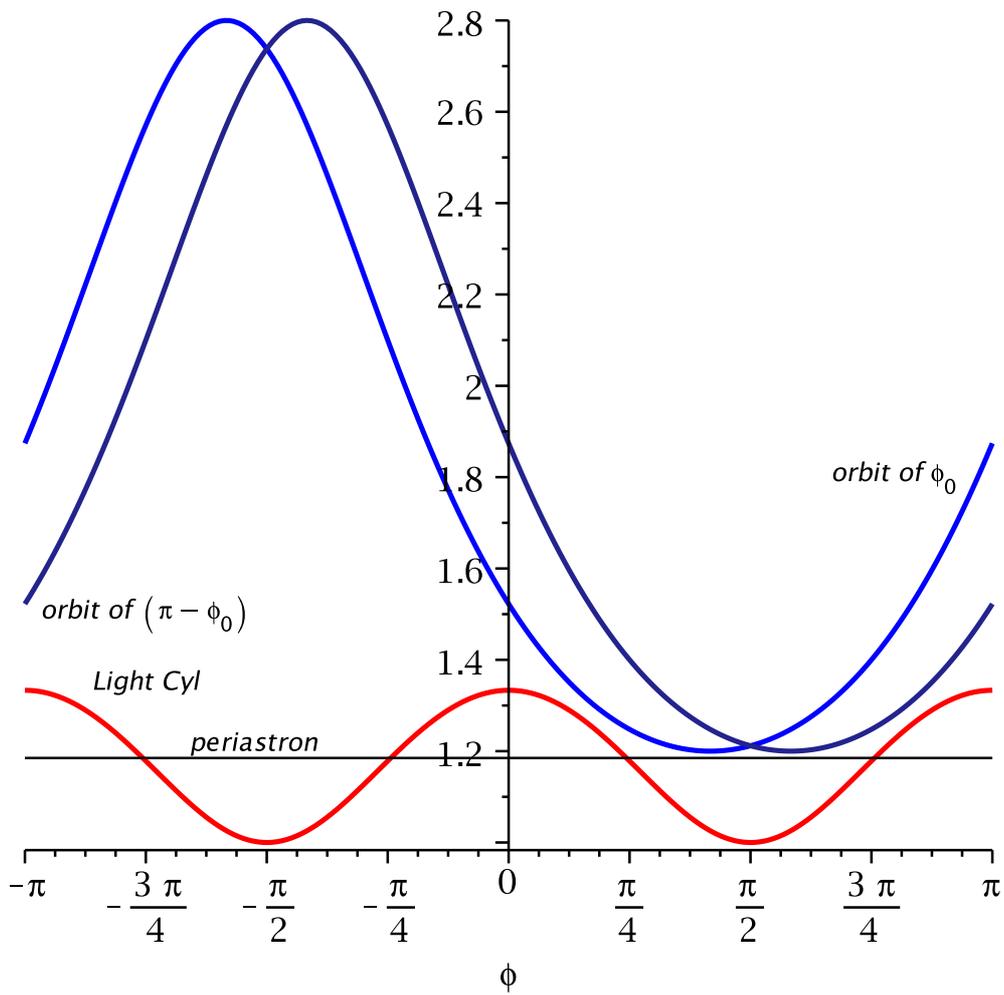}}
\caption{Plots as a function of $\phi$ (figure \ref{deux_ellipses}) of the distance to the star on orbits of periastron phases
$\phi_0$ and $\pi - \phi_0$, both entirely out of the light cylinder. The distance of the latter to
the star is  shown as well as the periastron distance. }
\label{outofCylLum}
\end{figure}

\newpage
\appendix

\section{Sufficient condition for active orbits}
\label{appsurpisurdeux}

The symmetries apparent in figure \ref{deux_ellipses} show that if an orbit of periapsis
phase $\phi_0$ in the quadrant
$[0, \pi/2]$ is entirely external to the light cylinder,
so are also the orbits of periapsis phases $(\pi - \phi_0)$, $(- \phi_0)$, $(\phi_0 - \pi)$.
We can then restrict ourselves to deriving the conditions for some orbit to be entirely out
of the light cylinder to the case when $0 < \phi_0\leq \pi/2$.
Figure \ref{outofCylLum} represents the orbital radii of
two orbits $\phi_0$ and $(\pi - \phi_0)$,  entirely external to the light cylinder.
The distance to the light cylinder in the orbital plane is also shown, as a function of the polar angle $\phi$.
Now, it can be understood from that figure that when the periapsis phase $\psi$  progressively increases from $\phi_0$ to $\pi - \phi_0$,
the orbital radius remains between a lower bound defined by
\begin{equation}
{\underline{r}}(\phi) = {\mathrm{Inf}} \left(
r_{\phi_{0}}(\phi), \, a(1-e), \, r_{(\pi - \phi_{0})}(\phi) \right)
\end{equation}
and an upper bound
\begin{equation}
{\overline{r}}(\phi) = {\mathrm{Max}} \left( r_{\phi_{0}}(\phi), \, a(1 +e), \, r_{(\pi - \phi_{0})}(\phi) \right).
\end{equation}
A glance at figure \ref{outofCylLum} reveals what a more formal proof could establish, namely that
the range of values allowed by these bounds for $r_\psi(\phi)$ when $\phi_0 < \psi <\pi - \phi_0$
is itself entirely out of the light cylinder. Since
$\pi/2$ is in this range of values,
it follows that if an orbit of phase $\phi_0$
is out of the light cylinder, so is also
the orbit of phase $\pi/2$ (which then is an on orbit). Conversely, if the orbit of phase $\pi/2$ is
entirely external to the light cylinder, then it is itself on.

\section{Sufficient conditions for both active and passive orbits}
\label{AppCNSonoff}

The maximum of $H(x)$, defined by Eq. (\ref{definHdex}), is reached at
\begin{equation} \label{xMdeeeti}
x_M= \frac{1}{4e}\, \left(-1+\sqrt{1+\frac{8e^2}{\sin^2 i}}\right).
\end{equation}
It is in the interval $[0, 1]$ provided $x_M \leq 1$. Otherwise, the maximum is reached 
in this interval at $x=1$, 
which occurs when
\begin{equation} \label{ineq_condition_1_i_petit}
i < i_{lim}(e) 
= \arccos\left(\sqrt{\frac{e+1}{2e+1}}\right)\, .
\end{equation}
The maximum value of $H$ on $[0,1]$ is in this case given by $H(1)= (1+e) \cos i$. It is lower than
the right-hand side of equation (\ref{ineq_condition_2_ellipses_phi0}) provided $R_{lc} < a (1-e)$. This
is part of the inequalities in Eq. (\ref{eq_geometry_condition1}) however, that have already been imposed
to grant that some orbits are off .
Thus, when Eq. (\ref{ineq_condition_1_i_petit}) applies,
no extra condition has to be imposed for on orbits to exist.

When $x_M \leq 1$, that is when $i$ is higher than the limit $i_{lim}(e)$
defined by Eq. (\ref{ineq_condition_1_i_petit}),
the condition for on orbits to exist can be written as
\begin{equation}\label{ineq_condition_2_final}
(1 + ex_M) \left(1 - x_M^2 \sin^2 i\right)^{1/2}  < \frac{a(1-e^2)\cos i}{R_{lc}}.
\end{equation}
For a given eccentricity in the range allowed by Eq. (\ref{ineq_condition1bis}),
the inequality in Eq. (\ref{ineq_condition_2_final}) is a condition imposed upon the inclination $i$.
This is complicated because $x_M$ depends on $i$,
albeit monotonically, as can be seen from Eq. (\ref{xMdeeeti}).
The equation $H'(x_M) = 0$, which defines $x_M$, writes:
\begin{equation}\label{equaderivenxMnul}
2 e x_M^2 \sin^2 i + x_M \sin^2 i - e = 0.
\end{equation}
It can be used to express $\sin^2 i$ and $\cos^2 i$ in terms of $x_M$ and $e$
and the result inserted into Eq. (\ref{ineq_condition_2_final}), which then becomes
an inequality constraining the values of $x_M$:
\begin{equation}\label{condisurxM}
x_M \ \left(\frac{1 + e x_M}{1 + e} \right)^3 \, \leq \, 
\left(\frac{a (1 -e)}{R_{lc} }\right)^2 \ \left(\frac{2 e x_M^2 + x_M - e}{1 +e}\right).
\end{equation}
The functions of $x_M$ that appear on the left and in the parenthesis on the right of Eq. (\ref{condisurxM})
both equal unity at $x_M = 1$, where they have the same derivative. The function on the right changes sign
between $x_M = 0$ and 1 at some value $x_{M0}$, which happens to be 
given by Eq. (\ref{xMdeeeti}) for $i = \pi/2$.
Both sides of the relation in Eq. (\ref{condisurxM}) are increasing functions of $x_M$. Due
to the constraints in Eq. (\ref{eq_geometry_condition1}), the factor in the first parenthesis on
the right of Eq. (\ref{condisurxM}) exceeds  unity, reaching it
only when $e = e_{max}$. In that particular case the inequality in Eq. (\ref{condisurxM}) is only
satisfied as an equality at $x_M = 1$. For values of $a (1 -e)/R_{lc}$ strictly exceeding unity,
the two curves representing the functions
on the left and on the right of Eq. (\ref{condisurxM}) intersect at some $x_{Mmin}$,
which is between $x_{M0}$ and 1. The value of $x_{Mmin}$ depends on $a/R_{lc}$. For a given object,
$a/R_{lc}$ is a function of $e$ given by Eq. (\ref{eq_a_de_C_e}) however.
The inequality in Eq. (\ref{condisurxM}) is satisfied for $x_M \geq x_{Mmin}$. The lower bound $x_{Mmin}$
corresponds for a given object and a given eccentricity $e$ to an upper bound to $i$, $i_ {max}(e)$:
\begin{equation}
i < i_ {max}(e).
\end{equation}
The variation of $i_{max}(e)$ with $e$
for a given object of known periapsis precession period
must be computed numerically.
To summarise, when $x_M < 1$, that is when $i > i_{lim}(e)$,
both on and off orbit can only exist, accounting for
the constraints (\ref{ineq_condition1bis}),
when $i_{min}(e) < i < i_{max}(e)$ and when at the same time $e < e_{max}$.

%
%


\begin{thebibliography}{23}
\expandafter\ifx\csname natexlab\endcsname\relax\def\natexlab#1{#1}\fi

\bibitem[{{Bailey} {et~al.}(1992){Bailey}, {Chambers}, \& {Hahn}}]{Bailey_1992}
{Bailey}, M.~E., {Chambers}, J.~E., \& {Hahn}, G. 1992, Astronomy and
  Astrophysics, 257, 315

\bibitem[{{Bonfond}(2010)}]{Bonfond_2010}
{Bonfond}, B. 2010, Journal of Geophysical Research (Space Physics), 115, 9217

\bibitem[{{Camilo} {et~al.}(2011){Camilo}, {Ransom}, {Chatterjee}, {Johnston},
  \& {Demorest}}]{Camilo_2011}
{Camilo}, F., {Ransom}, S.~M., {Chatterjee}, S., {Johnston}, S., \& {Demorest},
  P. 2011, ArXiv e-prints

\bibitem[{{Chan{\'e}} {et~al.}(2012){Chan{\'e}}, {Saur}, {Neubauer}, {Raeder},
  \& {Poedts}}]{Chane_2012}
{Chan{\'e}}, E., {Saur}, J., {Neubauer}, F.~M., {Raeder}, J., \& {Poedts}, S.
  2012, Journal of Geophysical Research (Space Physics), 117, 9217

\bibitem[{{Cook}(1973)}]{Cook_1973}
{Cook}, A.~H. 1973, {Physics of the earth and planets}

\bibitem[{{Espinoza} {et~al.}(2011){Espinoza}, {Lyne}, {Stappers}, \&
  {Kramer}}]{Espinoza_2011}
{Espinoza}, C.~M., {Lyne}, A.~G., {Stappers}, B.~W., \& {Kramer}, M. 2011,
  \mnras, 414, 1679

\bibitem[{{Guinier} \& {Jullien}(1987)}]{Guinier_1987}
{Guinier}, A. \& {Jullien}, R. 1987, {La mati\`ere \`a l'\'etat solide}

\bibitem[{{Helfand} {et~al.}(2006){Helfand}, {Becker}, {White}, {Fallon}, \&
  {Tuttle}}]{Helfand_2006}
{Helfand}, D.~J., {Becker}, R.~H., {White}, R.~L., {Fallon}, A., \& {Tuttle},
  S. 2006, \aj, 131, 2525

\bibitem[{{Hess} {et~al.}(2010){Hess}, {P{\'e}tin}, {Zarka}, {Bonfond}, \&
  {Cecconi}}]{Hess_2010_serpe}
{Hess}, S.~L.~G., {P{\'e}tin}, A., {Zarka}, P., {Bonfond}, B., \& {Cecconi}, B.
  2010, Planetary and Space Science, 58, 1188

\bibitem[{{Kramer} {et~al.}(2006){Kramer}, {Lyne}, {O'Brien}, {Jordan}, \&
  {Lorimer}}]{Kramer_2006}
{Kramer}, M., {Lyne}, A.~G., {O'Brien}, J.~T., {Jordan}, C.~A., \& {Lorimer},
  D.~R. 2006, Science, 312, 549

\bibitem[{{Lorimer}(2007)}]{Lorimer_2007}
{Lorimer}, D. 2007, in Chandra Proposal, 2411

\bibitem[{{Lorimer} \& {Kramer}(2012)}]{Lorimer_2012}
{Lorimer}, D.~R. \& {Kramer}, M. 2012, {Handbook of Pulsar Astronomy}

\bibitem[{{Luo} \& {Melrose}(2007)}]{Luo_2007}
{Luo}, Q. \& {Melrose}, D. 2007, \mnras, 378, 1481

\bibitem[{{Mottez}(2011)}]{Mottez_2011_Graz}
{Mottez}, F. 2011, in Proceedings of the 7th International Workshop on
  Planetary, Solar and Heliospheric Radio Emissions held at Graz, Austria, ed.
  H.~O. Rucker, W.~S. Kurth, P.~Louarn, \& G.~Fischer, 315--+

\bibitem[{{Mottez} {et~al.}(2013)}]{Mottez_1931p24_1}
Paper I, {Mottez}, F. \& {Bonazzola}, S., \& {Heyvaerts}, J. 2011{\natexlab{b}}, Astronomy and Astrophysics,  THIS ISSUE, TO BE COMPLETED

\bibitem[{{Mottez} \& {Heyvaerts}(2011{\natexlab{a}})}]{Mottez_2011_AWO}
{Mottez}, F. \& {Heyvaerts}, J. 2011{\natexlab{a}}, Astronomy and Astrophysics,
  532, A22+

\bibitem[{{Mottez} \& {Heyvaerts}(2011{\natexlab{b}})}]{Mottez_2011_AWW}
{Mottez}, F. \& {Heyvaerts}, J. 2011{\natexlab{b}}, Astronomy and Astrophysics,
  532, A21+

\bibitem[{{Neubauer}(1980)}]{Neubauer_1980}
{Neubauer}, F.~M. 1980, Journal of Geophysical Research (Space Physics), 85,
  1171

\bibitem[{{Pontius} \& {Hill}(2006)}]{Pontius_2006}
{Pontius}, D.~H. \& {Hill}, T.~W. 2006, Journal of Geophysical Research (Space
  Physics), 111, 9214

\bibitem[{{Preusse} {et~al.}(2006){Preusse}, {Kopp}, {B{\"u}chner}, \&
  {Motschmann}}]{Preusse_2006}
{Preusse}, S., {Kopp}, A., {B{\"u}chner}, J., \& {Motschmann}, U. 2006,
  Astronomy and Astrophysics, 460, 317

\bibitem[{{Rea} {et~al.}(2008){Rea}, {Kramer}, {Stella}, {Jonker}, {Bassa},
  {Groot}, {Israel}, {M{\'e}ndez}, {Possenti}, \& {Lyne}}]{Rea_2008}
{Rea}, N., {Kramer}, M., {Stella}, L., {et~al.} 2008, Monthly Notices of the
  Royal Astronomical Society, 391, 663

\bibitem[{{Shannon} {et~al.}(2013){Shannon}, {Cordes}, {Metcalfe}, {Lazio},
  {Cognard}, {Desvignes}, {Janssen}, {Jessner}, {Kramer}, {Lazaridis},
  {Purver}, {Stappers}, \& {Theureau}}]{Shannon_2013}
{Shannon}, R.~M., {Cordes}, J.~M., {Metcalfe}, T.~S., {et~al.} 2013, ArXiv
  e-prints

\bibitem[{{Willes} \& {Wu}(2005)}]{Willes_2005}
{Willes}, A.~J. \& {Wu}, K. 2005, Astronomy and Astrophysics, 432, 1091

\end{thebibliography}

\end{document}